\documentstyle[preprint,pra,aps]{revtex}
\begin{document}
\tighten
\title{Phonon exchange in dilute Fermi-Bose mixtures:
tailoring the Fermi-Fermi interaction.}
\author{M.J. Bijlsma, B.A. Heringa, and H.T.C. Stoof \\
Institute for Theoretical Physics, University of Utrecht, \\
Princetonplein 5, 3584 CC Utrecht, The Netherlands}
\maketitle
\begin{abstract}
We consider a mixture of a single-component Bose gas and a 
two-component Fermi gas at 
temperatures where the Bose gas is almost fully condensed.
In such a mixture, two fermionic atoms can interact with 
each other by exchanging a phonon that propagates through 
the Bose condensate. We calculate the interaction potential due
to this mechanism and determine the effective $s$-wave 
scattering length for two fermions that  
interact, both directly by the interatomic potentials, 
as well as by the above mentioned exchange mechanism.
We find that the effective scattering length is quite sensitive
to changes in the condensate density and becomes strongly
energy dependent. In addition, we consider the mechanical stability of 
these mixtures and also calculate the dispersion and the damping of 
the various collisionless collective modes of the gas.        
\end{abstract}
\section{INTRODUCTION}
The experimental realization of Bose-Einstein condensation in trapped 
atomic Bose gases \cite{anderson,bradley,davis} 
has stimulated both theoretical as well as experimental 
efforts to investigate new and interesting physics in 
ultracold atomic gases. In particular, 
it has spurred a lot of interest in achieving a BCS
transition in trapped fermionic gases, also because 
a spin-polarized gas of atomic $^6$Li 
has theoretically been shown to 
undergo a transition into a superfluid state at a critical temperature
that is comparable to those realized in the experiments
with trapped Bose gases \cite{marianne,modawi}.
This relatively high critical temperature 
is due to it's anomalously large and negative scattering length
of $-2160$ a$_{0}$ \cite{abraham}. 
To reach the temperature regime
where the quantum degeneracy of the atomic gas becomes important,
evaporative cooling is used. This technique, however, requires
a fast thermalization rate and therefore a high collision frequency.
For doubly spin-polarized Fermi gases this is not possible due to
the exclusion principle, and such a high collision frequency must be obtained by 
using a mixture of either Fermi-Bose gases \cite{eddy,you} or Fermi-Fermi gases 
\cite{stoof96}. Experimentally, this method of sympathetic
cooling has already been used to produce the first two-component Bose condensate 
\cite{myatt97} and has recently also been reported to be succesful for
a fermionic two-component $^{40}$K mixture \cite{demarco2}.

Mixtures of dilute atomic gases are, however, interesting in their own right, 
both from an experimental as well as from a theoretical viewpoint. 
Indeed, experimental work on the static and dynamic properties of a binary 
mixture of Bose gases has included the study of mean-field effects \cite{eric1}, 
relative-phase coherence \cite{hall1}, 
the dynamics of component separation \cite{hall2}, and Rabi oscillations 
\cite{eric2}. Furthermore, metastable 
states \cite{miesner} and quantum tunneling effects \cite{ketterle} have been 
observed in spinor Bose-Einstein condensates, and most 
recently a convenient method for the creation of topological excitations in 
two-component Bose-condensed gases has been suggested \cite{williams} and 
succesfully carried out experimentally \cite{matthews}.
Theoretical work on trapped binary Bose condensed gases, has for example 
included the stability and static properties \cite{ho2,green,ohberg}, the 
dynamics of the relative phase \cite{castin} and the collective modes 
\cite{busch}. Also for spinor Bose condensates in optical traps, the phase 
diagram and collective modes have been considered \cite{ho1,law}.
In the case of Fermi-Bose mixtures, 
the density profiles for gases 
confined in a harmonic trap have been studied at nonzero temperatures
in a Thomas-Fermi approximation \cite{molmer98,vichi98,amoruso98}. 
In all these cases the theory is based on a mean-field treatment 
of the interactions and neglects the effect of fluctuations. 

Inspired by the well-known physics of $^3$He-$^4$He mixtures \cite{helium},
we here go beyond mean-field theory and
study the effective interaction between two fermions
due to density fluctuations in a Bose condensate.
In particular, we calculate
the resulting effective interatomic $s$-wave scattering length for two fermions 
in different hyperfine states with the aim of manipulating
it in such a way that a BCS transition becomes feasible. The two
hyperfine states form an effective spin $1/2$ system,
and therefore the mixture of fermions in two 
different hyperfine states can be treated as 
a Fermi gas with spin $1/2$.
We mostly take the populations of the two spin levels to be equal
because, if a BCS transition can be achieved at all, this will be
the optimal situation. 
Moreover, we also consider the system to be homogeneous, for
the following reason:
If the mixture would be trapped in an
isotropic harmonic potential with trapping 
frequency $\omega$, a measure for the 
overlap of the Bose condensate
with the fermionic cloud 
is given by the ratio of 
the zero-temperature Thomas-Fermi radii for an interacting
Bose condensed gas and that of an ideal Fermi gas.
The former is equal to $l (15 N_{B} a_{B} / l)^{1/5}$ \cite{pethick},
where $l = \sqrt{\hbar / m \omega}$ is the harmonic oscillator length,
$a_{B}$ is the bosonic scattering length, and $N_{B}$ is the number 
of bosons. 
The latter equals $l (48 N_{F})^{1/6}$ \cite{dan}, where $N_{F}$ is the number 
of fermions.
For typical experimental parameters and the desirable large numbers of bosonic 
and fermionic atoms, this overlap
of the two clouds is rather small. Therefore, 
in order to maximize the effect of the Bose condensate
on the interaction between the fermions in the mixture, 
we consider only a spatially homogenous system. 
Although all the experiments with Bose condensed gases have been performed with 
harmonic oscillator traps up to now, this is not an unrealistic suggestion since 
it is certainly possible to create an external trapping potential
that is more or less a rectangular box \cite{communication}. 

Physically, the effect of the Bose condensate on the fermion-fermion
interaction is due to the exchange of phonons that propagate in
the Bose-condensed gas. In order to calculate the effect 
on the interatomic interaction 
quantitatively, we thus, have to accurately know the density-density correlation 
function in the Bose gas. 
This is straightforward in the Bogoliubov approximation, whose validity is 
well-established at such low temperatures that the Bose gas is essentially 
fully condensed.
In this manner we obtain the actual interaction potential from which
we then extract the interatomic scattering length. 
In addition to the effective fermion-fermion interaction, we 
consider the important question of the 
stability of the three-component system 
against demixing of the various Bose and Fermi components of the gas. We also 
calculate the excitation spectrum for the Fermi-Bose mixture far below
the Bose-Einstein transition temperature, where the dynamics of the gas is
in the collisionless regime. Theoretically, this amounts to doing a so-called 
RPA (Random Phase Approximation)  
calculation for this mixture. As a result we find not only
the eigenmodes of the gas, but also the Landau damping
of these modes due to the imaginary part of the `RPA-bubble' diagram.
Note that from a fundamental point of view the stability and the excitation 
spectrum are strongly related, because a signature of the demixing-instability 
is the occurrence of a mode with a purely imaginary frequency. 

The paper is organized as follows. 
In Sec. \ref{sec2a} we first calculate the effective interaction
and scattering length of two fermions in the presence of a Bose condensate. 
In particular, we show that
for realistic conditions, the scattering length strongly depends
on the collision energy of the atoms, which is 
important when considering the prospects of a BCS transition in a 
spin-polarized potassium gas.  
In Sec. \ref{sec2b} 
we then consider the stability of the three-component system
and show that in general this does not lead to very stringent constraints
on the densities. Finally, in Sec. \ref{sec2c},
we calculate the long-wavelength collective mode 
spectrum and damping for the Fermi-Bose mixture.
We end with a summary in Sec.~\ref{sec2d}.

\section{EFFECTIVE INTERACTION}
\label{sec2a}
In this section, we calculate the effective interaction
and scattering length of two fermions. The  
calculation is performed by means of functional methods \cite{henk}, because 
even in the presence of a Bose condensate, 
the RPA calculation for the collective modes 
can then be performed in a purely algebraic manner in Sec. \ref{sec2c} 
and avoids the complications of an explicit evaluation of Feynman diagrams. 
Moreover, it gives some more insight in the physical nature of the collisionless
collective modes. Of course, it is also possible to do the same calculation
in the operator formalism. If performed correctly, it gives identical results.

\subsection{Theory}
In accordance with the previous remarks, we thus start from the 
functional-integral 
expression for the grand-canonical partition function of the mixture. It reads 
\begin{eqnarray}
{\cal Z}_{\rm gr} = \int d[\phi^{*}]d[\phi]d[\psi^{*}]d[\psi]
\exp \left\{ - {1 \over \hbar} {\Big(} 
S_{B}[\phi^{*},\phi] + S_{F}[\psi^{*},\psi] + 
S_{I}[\phi^{*},\phi,\psi^{*},\psi] {\Big)} \right\} \; ,  
\end{eqnarray}
and consists of an integration over a complex field $\phi({\bf x},\tau)$, which 
is periodic on the imaginary-time interval $[0,\hbar\beta]$, and over the 
Grassmann fields $\psi_{\alpha}({\bf x},\tau)$, which are antiperiodic on this 
interval. Therefore, $\phi({\bf x},\tau)$ describes the Bose component of the 
mixture, whereas $\psi_{\alpha}({\bf x},\tau)$ is associated with the Fermi 
components. For the latter we actually need two fields, because the fermionic 
atoms can be in either one of the two hyperfine states $|\alpha\rangle$.
Furthermore, the total action of the mixture consists of a term for the Bose-gas
\begin{eqnarray}
\label{eq1}
S_{B}[\phi^{*},\phi] = 
\int_{0}^{\hbar\beta} d\tau  \int  d{\bf x} 
\left\{ \phi^{*}({\bf x},\tau)
\left( \hbar {\partial \over \partial \tau} -
{\hbar^2\mbox{\boldmath $\nabla$}^2 \over 2m_{B}} - \mu \right) 
\phi({\bf x},\tau) 
+ {T_{B} \over 2} |\phi({\bf x},\tau)|^{4} \right\} \; ,
\end{eqnarray}
a term for the Fermi-gas that accounts for the fact that the Pauli principle 
forbids $s$-wave scattering between fermionic atoms in the same hyperfine state
\begin{eqnarray}
\label{eq2}
S_{F}[\psi^{*},\psi] &=& \sum_{\alpha}  
\int_{0}^{\hbar\beta} d\tau  \int  d{\bf x}
\left\{ \psi^{*}_{\alpha}({\bf x},\tau)
\left( \hbar {\partial \over \partial \tau} - 
{\hbar^2\mbox{\boldmath $\nabla$}^2 \over 2m_{F}} - \mu_{\alpha} \right) 
\psi_{\alpha}({\bf x},\tau) \nonumber \right. \\
&& + \left. 
{T_{F} \over 2} |\psi_{\alpha}({\bf x},\tau)|^{2}  
|\psi_{-\alpha}({\bf x},\tau)|^{2} \raisebox{0.25in}{} \right\} \; ,
\end{eqnarray}
and a term describing the interaction between the three components of the 
Fermi-Bose mixture
\begin{eqnarray}
\label{eq3}
S_{I}[\phi^{*},\phi,\psi^{*},\psi] &=& \sum_{\alpha}
T_{\alpha} 
\int_{0}^{\hbar\beta} d\tau  \int  d{\bf x}
|\psi_{\alpha}({\bf x},\tau)|^{2}
|\phi({\bf x},\tau)|^{2} \; .
\end{eqnarray}
In these expressions we have introduced the two-body boson-boson
T(ransition) matrix element $T_{B} = 4 \pi \hbar^{2} a_{B} / m_{B}$,
the two-body fermion-fermion T-matrix element
$T_{F} = 4 \pi \hbar^{2} a_{F} / m_{F}$, 
and the two two-body T-matrix elements
$T_{\alpha} = 2 \pi \hbar^{2} a_{\alpha} / m_{R}$
that describe the interactions between a boson and a fermion in the spinstate 
$|\alpha\rangle$. Here $\alpha=\{ \uparrow,\downarrow \}$ denotes the 
hyperfine components of the
Fermi-gas in the effective spin language. 
In addition, $\mu$ denotes the chemical potential
of the Bose-gas and $\mu_{\alpha}$ denotes the chemical 
potentials of the two components of the Fermi gas.
Note that the latter do not need to be identical, because the Fermi gas is
in general not in equilibrium in spin space, due to the generally
slow relaxation rates between the hyperfine degrees of freedom.
Finally, the masses of the bosonic atoms and the fermionic atoms 
are denoted by $m_{B}$ and $m_{F}$, respectively.
The reduced mass is denoted by $m_{R}=m_{F}m_{B}/(m_{F}+m_{B})$.

Since we only consider the gas at such low temperatures
that the Bose gas is essentially fully Bose condensed,
we proceed by performing the usual Bogoliubov substitution
for the Bose-fields, i.e., $\phi=\sqrt{n_{B}}+\phi'$,
and neglect all terms of higher than second order in $\phi'$ or $\phi'^*$. Note 
that we have implicitly also neglected the depletion of the condensate and put 
the condensate density equal to the total density $n_B$ of the Bose gas. For a 
weakly-interacting gas with $\sqrt{n_Ba_B^3} \ll 1$ and at low temperatures, 
this is clearly justified. In this manner we end up with the Bogoliubov 
approximation 
$S_{B}[\phi^{*},\phi] \simeq \hbar\beta V T_B n_B^2/2 + S_{B}[\phi'^{*},\phi']$,
where $V$ is the total volume of the gas and the action for the fluctuations can 
be written as
\begin{eqnarray}
S_{B}[\phi'^{*},\phi'] &=& - {\hbar \over 2} 
\int_{0}^{\hbar\beta} d\tau d\tau'  \int  d{\bf x} d{\bf x'} 
\bbox{\phi}'^{\dagger}({\bf x},\tau) 
\cdot {\bf G}^{-1}({\bf x},\tau;{\bf x'},\tau') 
\cdot \bbox{\phi}'({\bf x'},\tau') \; ,
\end{eqnarray}
if we introduce the vector field 
\begin{eqnarray}
\bbox{\phi}'({\bf x},\tau) & = &
\left(
\begin{array}{c}
\phi'({\bf x},\tau) \\
\phi'^{*}({\bf x},\tau) 
\end{array}     
\right) \; ,
\end{eqnarray}
and the corresponding Green's function ${\bf G}({\bf x},\tau;{\bf x'},\tau')$ 
that obeys
\begin{eqnarray}
\label{gf}
- \hbar &&{\bf G}^{-1}({\bf x},\tau;{\bf x'},\tau')          \nonumber \\
&& = \left(
\begin{array}{cc}
\hbar\partial_\tau-\hbar^2\mbox{\boldmath $\nabla$}^2/2m+T_{B}n_{B} & 
Tn_{B} \\
Tn_{B} & 
-\hbar\partial_\tau-\hbar^2\mbox{\boldmath $\nabla$}^2/2m+T_{B}n_{B} \\
\end{array}
\right) 
\delta({\bf x-x'}) \delta(\tau-\tau')~.
\end{eqnarray}
Note that the linear terms in $\phi'$ and $\phi'^*$ have dropped out of the 
action because of the Hugenholtz-Pines relation
\begin{eqnarray}
\mu = T_{B} n_{B} + \sum_{\alpha} T_{\alpha} n_{\alpha} \; ,
\end{eqnarray}
which also incorporates the mean-field effects due to the nonzero spin densities 
$n_{\alpha}$ in the Fermi gas. The same relation is also used to eliminate the 
chemical potential from Eq.~(\ref{gf}).
In addition, the interaction term in the total action becomes in the Bogoliubov
approximation
\begin{eqnarray}		
S_{I}[\phi'^{*},\phi',\psi^{*},\psi] \simeq {1 \over 2}
\int_{0}^{\hbar\beta} d\tau  \int  d{\bf x} 
\left\{ {\bf J}^{\dagger}({\bf x},\tau) \cdot \bbox{\phi}'({\bf x},\tau) + 
\bbox{\phi}'^{\dagger}({\bf x},\tau) \cdot {\bf J}({\bf x},\tau) \right\} \; ,
\end{eqnarray}
with a `current source' defined by
\begin{eqnarray}
{\bf J}({\bf x},\tau) = 
\sqrt{n_{B}}  
\sum_{\alpha} T_{\alpha} 
|\psi_{\alpha}({\bf x},\tau)|^{2} \left(
\begin{array}{c}
1 \\
1
\end{array}     
\right) \; .
\end{eqnarray}

To include the effect of phonon exchange on the interaction between
two fermions, we now need to integrate over the Bose fields. This
is straightforward, since it only involves the evaluation of a gaussian 
integral. The result for the total effective fermion action is thus
\begin{eqnarray}
&& \hspace*{-0.2in}
S^{{\rm eff}}[\psi^{*},\psi] = S_{F}[\psi^{*},\psi] + {1 \over 2 \hbar} 
\int_{0}^{\hbar\beta} d\tau d\tau'  \int  d{\bf x}d{\bf x'}
{\bf J}^{\dagger}({\bf x},\tau) \cdot {\bf G}({\bf x},\tau;{\bf x'},\tau') 
\cdot {\bf J}({\bf x'},\tau')
\nonumber \\
&& \equiv
S_{F}[\psi^{*},\psi] + {1 \over 2} \sum_{\alpha,\alpha'}
\int_{0}^{\hbar\beta} d\tau d\tau'  \int  d{\bf x}d{\bf x'}
|\psi_{\alpha}({\bf x},\tau)|^{2}
V_{\alpha,\alpha'}({\bf x},\tau;{\bf x'},\tau') 
|\psi_{\alpha'}({\bf x}',\tau')|^{2} \; ,
\end{eqnarray}
where $V_{\alpha,\alpha'}({\bf x},\tau;{\bf x'},\tau')$ is the
effective interatomic potential due to phonon exchange.
At this point, it is important to note that to be able to use 
in Eqs.~(\ref{eq1}), (\ref{eq2}) and (\ref{eq3}) 
the two-body T-matrix elements, instead of
the real interatomic (singlet/triplet) potentials, we have
already integrated out fluctuations with momenta higher than a certain
cutoff $\hbar\Lambda$. Thus, in principle there is a cutoff on all the
momentum integration in the rest of this paper. We will comment
on the effects of this shortly. 
Hence, inverting Eq.~(\ref{gf}) by means of a Fourier transformation, we find 
that $V_{\alpha,\alpha'}({\bf x},\tau;{\bf x'},\tau')$ is given by
\begin{eqnarray}
\label{eq11}
&& \hspace*{-0.2in}
V_{\alpha,\alpha'}({\bf x},\tau;{\bf x'},\tau')                \nonumber \\
&& = - 2 T_{\alpha} T_{\alpha'} n_{B} 
\sum_{n} \int_{k \leq \Lambda} {d {\bf k} \over (2 \pi)^{3}} 
e^{i{\bf k \cdot (x-x')} - i \omega_{n}(\tau-\tau')}
\left[
{ \epsilon({\bf k}) \over
(\hbar \omega_{n})^{2} +\epsilon({\bf k})(\epsilon({\bf k}) + 2 T_{B} n_{B})} 
\right] \; ,
\end{eqnarray}
where $\epsilon({\bf k})=\hbar^{2} {\bf k}^{2} / 2 m$ and 
$\omega_{n}=2\pi n/\hbar\beta$ are the even Matsubara frequencies that account 
for the periodicity of the Bose field $\phi({\bf x},\tau)$ and therefore of the 
Green's function
${\bf G}({\bf x},\tau;{\bf x'},\tau')$. The phonon-exchange mechanism thus
also induces an interaction between fermions
with the same spins. Due to the Pauli-principle, however, this
interaction can again at best be of a $p$-wave nature, and is
in general negligible. As a result, we from now on only consider the 
contributions to the interaction potential 
between particles with opposite spin. 
We also neglect the frequency dependence in Eq.~(\ref{eq11}), and
consider only the static contribution. This implies that the relevant collision
energies of the fermions must be much less than the bosonic mean-field 
interaction. 

The resulting instantaneous potential is given by 
$V({\bf x},\tau;{\bf x}',\tau') 
                        \simeq V^{\rm eff}(|{\bf x-x'}|) \delta(\tau-\tau')$ 
with
\begin{eqnarray}
\label{eq12}
V^{\rm eff}(r)=-
{\hbar^{2} m_{B} a_{\uparrow} a_{\downarrow} 
                          \over 2 \pi m_{R}^{2} a_{B} \xi^{2} r}
\int_{0}^{\Lambda} \; dk 
{k \sin(k r) \over k^{2} + 1/\xi^{2}} \; .
\end{eqnarray}
Here we have defined the coherence length 
$\xi=1/\sqrt{16 \pi n_{B} a_{B}}$ in the
Bose-condensed gas.
The effective potential depends on three parameters,
the cutoff $\Lambda$, the coherence length $\xi$ and 
$m_{B} a_{\uparrow} a_{\downarrow} / m_{R}^{2} a_{B}$.
As we will show now, however, there is a separation of length
scales, and $1 / \Lambda$ is in general much smaller than
the other relevant length scale, the coherence length $\xi$.
This means that we can safely ignore the cutoff and 
take the limit $\Lambda \rightarrow \infty$. 
To estimate the magnitude of the cutoff, we make use of the fact that
it is determined by the requirement that 
the bare interatomic interaction has renormalized to the 
two-body scattering matrix. 
This renormalization of the bare interaction is described
by the Lippmann-Schwinger equation
\begin{eqnarray}
{1 \over T_{\Lambda}} & = &
{1 \over V_{0}} +
\int_{\Lambda < k \le \Lambda_{\rm a}} 
{d {\bf k} \over (2 \pi)^{3}}
{1 \over 2 \epsilon({\bf k})} \; .
\end{eqnarray}
Here $\Lambda_{\rm a}$ denotes the ultra-violet cutoff
provided by the interatomic potential, and $\Lambda$ is
the momentum scale up to which fluctuations in the
Bose gas are integrated over. The bare interaction
$V_{0}$ is chosen such, that in the limit $\Lambda=0$
the result does not depend on the high-momentum cutoff $\Lambda_{\rm a}$
\cite{bijlsma}. Requiring
the renormalized interaction to be
within approximately a fraction $x$ from the two-body transition matrix
leads to
\begin{eqnarray}
\Lambda \simeq \frac{x \pi}{2 a_{B}} \; .
\end{eqnarray}
If we compare this value for the cutoff with 
the coherence length $\xi$, the product
of the two is given by
\begin{eqnarray}
\xi \Lambda \simeq {x \pi \over 8 \sqrt{\pi n_{B} a_{B}^{3}}} \; .
\end{eqnarray}
Due to the presence of the factor $\sqrt{n_{B} a_{B}^{3}}$ in the denominator, 
this quantity is in general much larger than $1$, and we can safely take the 
limit $\Lambda \rightarrow \infty$. The effective interaction in this case 
becomes simply
\begin{eqnarray}
V^{\rm eff}(r)=-{\hbar^{2} 
m_{B} a_{\uparrow} a_{\downarrow} \over 4 m_{R}^{2} a_{B} \xi^{2} r}
e^{- r / \xi} \; ,
\end{eqnarray}
and has the form of a purely attractive Yukawa potential.

\subsection{Results}
With this interaction potential $V^{\rm eff}(r)$,
we want to determine the effective {\it s}-wave scattering 
length $a_{F}^{\rm eff}$ for two fermions interacting also
through the mechanism of phonon exchange. 
We can associate an {\it s}-wave scattering length
$a_{F}^{{\rm eff}}$ with the effective interaction $V^{\rm eff}(r)$ as follows
\begin{eqnarray}
a_{F}^{{\rm eff}}= - \lim_{k \downarrow 0} {\delta_{0}(k) \over k} \; .
\end{eqnarray}
Here $\delta_{0}(k)$ denotes the
phase shift of the partial wave with angular momentum $l=0$.
The phase shift $\delta_{0}(k)$
is defined in terms of the asymptotic form
\begin{eqnarray}
\lim_{r \rightarrow \infty} u_{0}(r;k) \rightarrow \sin(kr + \delta_{0}(k))
\end{eqnarray}
for the $l=0$ partial wave $u_{0}(r;k)$, that can be calculated from
the radial Schr\"odinger equation
\begin{eqnarray}
\left(- {d^{2} \over dr^{2}} + { m_F \over \hbar^{2}} V^{\rm eff}(r)  
      - k^2
\right) u_{0}(r;k) = 0~.
\end{eqnarray}
At this point, we have to keep in mind that there is already
a scattering length $a_{F}$ for the fermions
due to the interatomic potential. We can take
this interatomic scattering length
into account by imposing suitable boundary conditions on
the partial wave $u_{0}(r;k)$, such that we recover
$a_{F}$ if the interaction due to phonon exchange vanishes, 
i.e., if $V^{\rm eff}(r)=0$. The boundary conditions imposed
are such that the derivative and the magnitude 
of the wave function at $r=0$ are equal to those of the function  
$\sin[k(r-a_{F})]$. In the case of a positive scattering
length, an alternative procedure would be to add
a hard-core potential with a range $a_{F}$ to $V^{\rm eff}(r)$.
Both these procedures are justified because
the range of the interatomic potential is much
smaller than the range of the effective potential,
and we have checked that numerically they indeed yield the same results. 

The phase shifts
due to the effective potential $V^{\rm eff}(r)$ 
for mixtures of $^{40}$K and $^{6}$Li with $^{87}$Rb, 
are show in Figs.~\ref{fig1} and \ref{fig2}, respectively.
The interatomic scattering lengths for $^{40}$K-$^{40}$K, 
$^{6}$Li-$^{6}$Li and $^{87}$Rb-$^{87}$Rb collisions are taken to be 
$160$ a$_{0}$, $-2160$ a$_{0}$ and $109$ a$_{0}$, respectively 
\cite{abraham,demarco}. The scattering lengths for 
$^{40}$K and $^{6}$Li with $^{87}$Rb are, as far as we know,
not known and we have taken them to be equal to $100$ a$_{0}$, which
is presumably a typical value.
In Fig.~\ref{fig3} also the phase shifts for a mixture of $^{40}$K and $^{39}$K 
are shown, where the scattering length for 
$^{39}$K-$^{39}$K and $^{40}$K-$^{39}$K
collisions has been taken to be $5$ a$_{0}$ and 
$1000$ a$_{0}$, respectively \cite{demarco}. 
The various lines corresponds to different values of the condensate density.
It is clear from these figures that 
for a given mixture, the phase shifts as a function of momentum asymptotically 
all have the same slope independent of the value of the 
condensate density. This
slope corresponds to the interatomic scattering length $a_F$. However,
at long wavelengths the phase shifts can be
significantly different from  the asymptotic limit, and 
can depend strongly on the collisional energy. For a 
mixture of $^{40}$K-$^{87}$Rb this is certainly the case. 
The effective interaction can even become attractive, instead of
repulsive, around the Fermi momentum, which opens the possibility for a
BCS transition to a superfluid phase just like the exchange of phonons in the 
lattice leads to superconductivity in metals \cite{BCS}. 
For a mixture of $^{6}$Li-$^{87}$Rb the effects are less
pronounced, and amount to a slight enhancement of the
already very large and negative scattering length.
The case of $^{40}$K-$^{39}$K may be exciting, because
of the possibility of a rather large scattering length for 
$^{40}$K-$^{39}$K collisions, which can be of the order of $1000$ a$_{0}$
\cite{demarco}. If this value is correct, there are resonances in the
effective scattering length as shown in Fig.~\ref{fig4}.
This strongly resembles the existence of Feshbach resonances 
in the scattering length as a function of the applied bias magnetic field
\cite{tiesinga}.

\section{STABILITY}
\label{sec2b}
To be able to interact, the Bose and the two-component 
Fermi gases have to be overlapping. 
Therefore, we want to consider the stability of the gas against
demixing of the Fermi and Bose components. 
To determine the stability of the mixture, we consider
the matrix of second-order derivatives with respect to the densities of 
the free energy $F=F_{F}+F_{B}+F_{I}$, which consists of a fermion term $F_{F}$, 
a boson term $F_{B}$ and an interaction term $F_{I}$. 
For the mixture to be stable we have to require that
all eigenvalues of this matrix are larger than zero,
i.e., that the free-energy surface is convex.
At the low temperatures of interest to us, the free-energy density  
is equal to the average energy density $\langle E \rangle/V$ and we get

\begin{eqnarray}
\langle E \rangle/ V & = & \langle E_{F} \rangle/V 
+ \langle E_{B} \rangle/V + \langle E_{I} \rangle/V
\nonumber \\
& = & {3 \over 10} (6 \pi^{2})^{2/3} (n_{\uparrow}^{5/3} 
+ n_{\downarrow}^{5/3}) {\hbar^{2} \over m_{F}}
+ T^{\rm eff}_{F} n_{\uparrow} n_{\downarrow} 
+ T_{B} {n_{B}^{2} \over 2} 
+ (T_{\uparrow}n_{\uparrow}  
+ T_{\downarrow} n_{\downarrow}) n_{B} \; .
\end{eqnarray}
Here, the first term is the kinetic energy of the two-component Fermi gas, 
the second and third terms are the interaction energies of the 
individual Fermi and the Bose gases, 
respectively, and the fourth term is due to the interaction between these gases.
We have neglected the kinetic energy of the Bose gas 
because of the low temperatures and densities.
Note that the effective fermion-fermion T-matrix element
$T^{\rm eff}_{F}=4 \pi \hbar^{2} a^{\rm eff}_{F} / m_{F}$ has to be used, 
that includes the effect of phonon exchange.
This is the case because, as long as the Fermi energy is much less than the 
average mean-field energy of the Bose condensate, we have to include 
the effect of fluctuations, which change the static properties
of the mixture by renormalizing the fermion-fermion interaction
as we have just discussed. 

The matrix of second-order derivatives is given by
\begin{eqnarray}
\left(
\begin{array}{ccc}
{\partial^{2} F \over \partial n_{\uparrow} \partial n_{\uparrow}} &
{\partial^{2} F \over \partial n_{\uparrow} \partial n_{\downarrow}} &
{\partial^{2} F \over \partial n_{\uparrow} \partial n_{B}} \\
{\partial^{2} F \over \partial n_{\downarrow} \partial n_{\uparrow}} &
{\partial^{2} F \over \partial n_{\downarrow} \partial n_{\downarrow}} &
{\partial^{2} F \over \partial n_{\downarrow} \partial n_{B}} \\
{\partial^{2} F \over \partial n_{B} \partial n_{\uparrow}} &
{\partial^{2} F \over \partial n_{B} \partial n_{\downarrow}} &
{\partial^{2} F \over \partial n_{B} \partial n_{B}}
\end{array}
\right) \; . \nonumber 
\end{eqnarray}
Assuming that we start in the stable part of the phase diagram,
the onset of an instability is signaled by the point
where its determinant becomes equal to zero.
Therefore, requiring the determinant of this matrix to be larger than
zero is a sufficient condition for stability. It reads
\begin{eqnarray}
\label{eq21}
4 \left(
{\pi \over k_{\uparrow} a_{F}^{\rm eff}}
\right)
\left(
{\pi \over k_{\downarrow} a_{F}^{\rm eff}}
\right)
& - & {2 \over a_{B} a_{F}^{\rm eff}}
{m_{B} m_{F} \over m_{R}^{2}}
\left[
\left(
{\pi \over k_{\downarrow} a_{F}^{\rm eff}}
\right)
a_{\uparrow}^{2} +
\left(
{\pi \over k_{\uparrow} a_{F}^{\rm eff}}
\right)
a_{\downarrow}^{2}
\right]
\nonumber \\
& + & 8 {a_{\downarrow} a_{\uparrow} \over a_{B} a_{F}^{\rm eff}}
{m_{B} m_{F} \over m_{R}^{2}} - 16 \ge 0 \; ,
\end{eqnarray}
where $k_{\alpha}=(6 \pi^{2} n_{\alpha})^{1/3}$ denotes the Fermi momentum
associated with the spin state $|\alpha\rangle$. 
Note that $a_F^{{\rm eff}}$ is a function of the condensate density $n_{B}$. In 
evaluating the derivatives of the free energy with respect to the density of the 
Bose gas, we have not taken this implicite dependence into account, which is 
sufficiently accurate for our purposes as long as we are not too close to a 
resonance. Near a resonance a more involved treatment is necessary, also because 
the phaseshift is then very strongly momentum dependent.

The surface in the $n_{\alpha}-n_{B}$ volume where the
equality sign holds is called the spinodal surface. 
The spinodal surface divides the phase space into a region where
the mixture is (meta)stable, and one where it is unstable and 
separates into distinct phases in the
stable part of the phase space.
If the effect of phonon exchange is optimized by 
putting $n_{\uparrow}=n_{\downarrow}$,
we have $k_{\downarrow}=k_{\uparrow} \equiv k_{F}$, and
Eq.~(\ref{eq21}) becomes
\begin{eqnarray}
4 \left(
{\pi \over k_{F} a_{F}^{\rm eff}}
- 2 \right)
\left(
{\pi \over k_{F} a_{F}^{\rm eff}}
+ 2 \right)
& - & {(a_{\uparrow} + a_{\downarrow})^{2} \over a_{B} a_{F}^{\rm eff}}
{m_{B} m_{F} \over m_{R}^{2}}
\left(
{\pi \over k_{F} a_{F}^{\rm eff}} - 2 
\right) 
\nonumber \\
& - &
{(a_{\uparrow} 
- a_{\downarrow})^{2} \over a_{B} a_{F}^{\rm eff}}
{m_{B} m_{F} \over m_{R}^{2}}
\left(
{\pi \over k_{F} a_{F}^{\rm eff}} + 2 
\right) \ge 0 \; .
\end{eqnarray}
If we also put $a_{\uparrow} = a_{\downarrow}=a_{FB}$, the condition
further simplyfies to
\begin{eqnarray}
\label{eq24}
 \left(
{\pi \over k_{F} a_{F}^{\rm eff}}
- 2 \right)
\left(
{\pi \over k_{F} a_{F}^{\rm eff}}
+ 2 -
{a_{FB}^{2} \over a_{B} a_{F}^{\rm eff}}
{m_{B} m_{F} \over m_{R}^{2}}
\right) \ge 0 \; .
\end{eqnarray}
The first factor in the left-hand side of Eq.~(\ref{eq24}) corresponds to the 
demixing of the two fermion components of the gas \cite{marianne}, whereas the
second factor represents the demixing of the Bose-Einstein condensate
and the fermion components \cite{luciano}. Note that it depends on the
various masses and scattering lengths involved, which of the 
two instabilities occurs first. For equal scattering lengths, i.e., 
$a_{F}^{\rm eff} = a_{B} = a_{\uparrow} = a_{\downarrow} \equiv a$,
the demixing of the Bose and Fermi gases always occurs first,
and we reproduce the result of van 
Leeuwen and Cohen \cite{vanleeuwen},
\begin{eqnarray}
{\pi \over k_{F} a} \ge {m_{B} \over m_{F}} + {m_{F} \over m_{B}} \; .
\end{eqnarray}

In our numerical calculations, the scattering
lengths $a_{\uparrow}$ and  $a_{\downarrow}$ have always been
taken equal to each other and Eq.~(\ref{eq24}) applies. In the case of a 
mixture of $^{40}$K and $^{87}$Rb this condition 
roughly leads for the total fermion density $n_F$ only to the restriction 
$n_{F} < 10^{18}$ cm$^{-3}$. 
For a mixture of $^{6}$Li and $^{87}$Rb the condition on the total density   
becomes $n_{F} < 10^{15}$ cm$^{-3}$, and for a 
mixture of $^{40}$K and $^{39}$K it reads $n_{F} < 10^{10}$ cm$^{-3}$.
The latter condition seems to be quite restrictive. However,
if we take instead of $a_{B}=5$ a$_{0}$ a different value
that is within the present uncertainty for this scattering length, 
i.e., $a_{B}=25$ a$_{0}$,
the condition for a mixture of $^{40}$K and $^{39}$K 
becomes only $n_{F} < 10^{12}$ cm$^{-3}$, which is much more
favorable when one considers the prospects of achieving
a BCS transition in this case. The reason is that in BCS
theory, the critical temperature is given by 
$T_{c} =
(8 \epsilon_{F} e^{\gamma-2}/ k_{B}\pi) 
                    \exp \left\{- \pi {\rm cot}(\delta_0(k_{F}))/2 \right\}$,
with $\gamma$ Euler's constant. From this expression it follows that if the 
densities are low,
the critical temperature for the BCS transition is also very low.
Therefore, if the results of 
Sec.~\ref{sec2a} for a $^{40}$K and $^{39}$K mixture
are to be of use, as far as achieving a BCS transition is concerned, it is 
crucial that relatively high densities are realizable.  

\section{DISPERSION OF COLLECTIVE EXCITATIONS}
\label{sec2c}
We next want to consider the excitation spectrum
of the gas. The collective excitations 
of the mixed gas are coupled modes 
of the fermionic spin densities, and the Bose condensate. 
It is therefore convenient to perform a Hubbard-Stratonovich
transformation to these fermionic densities
\cite{kleinert}. This amounts 
to introducing two real auxiliary fields 
$\rho_{\uparrow}({\bf x},\tau)$ and $\rho_{\downarrow}({\bf x},\tau)$,
by rewriting in the integrant of the functional integral for the partition 
function ${\cal Z}_{\rm gr}$, the factor due to the fermion-fermion interaction 
as,
\begin{eqnarray}
&& 
\exp \left\{ - {T_{F}\over \hbar} 
\int_{0}^{\hbar\beta}  d\tau  \int  d{\bf x}
|\psi_{\uparrow}({\bf x},\tau)|^{2}  
|\psi_{\downarrow}({\bf x},\tau)|^{2} \right\} 
= \int d[\rho_{\uparrow}] d[\rho_{\downarrow}]
\nonumber \\
&& \times
\exp \left\{ {T_{F} \over \hbar} 
\int_{0}^{\hbar\beta}  d\tau  \int  d{\bf x}
\left[ 
\rho_{\uparrow}({\bf x},\tau)
\rho_{\downarrow}({\bf x},\tau)  
- \rho_{\uparrow}({\bf x},\tau) 
|\psi_{\downarrow}({\bf x},\tau)|^{2}
- |\psi_{\uparrow}({\bf x},\tau)|^{2}
\rho_{\downarrow}({\bf x},\tau) \right]  
\right\} \; .
\end{eqnarray}
For reasons that become clear shortly, we denote the resulting action 
for the fermions by $S^{H}_{F}[\psi^{*},\psi]$.
It is quadratic in the fermion fields, and reads

\begin{eqnarray}
\label{eq27}
S^{H}_{F}[\psi^{*},\psi] = \sum_{\alpha}  
\int_{0}^{\hbar\beta} d\tau  \int  d{\bf x}
\left\{ \psi^{*}_{\alpha}({\bf x},\tau)
\left[ \hbar {\partial \over \partial \tau} - 
{\hbar^2\mbox{\boldmath $\nabla$}^2
\over 2m_{F}} \right. \right. &-& \mu_{\alpha}
+ T_{F} \rho_{-\alpha}({\bf x},\tau) 
\nonumber \\ 
&+& \left. \left. \raisebox{0.25in}{}
T_{\alpha} |\phi({\bf x},\tau)|^{2} \right] 
\psi_{\alpha}({\bf x},\tau) \right\} \; ,
\end{eqnarray}
Upon integrating over the fermionic fields, the grand-canonical
partition function describing the gas is equal to a functional integral over the 
Bose field and the 
density fields only, with an effective action that reads
\begin{eqnarray}
\label{eq17}
S^{\rm eff}[\rho_{\uparrow},\rho_{\downarrow},\phi^{*},\phi]
= &-& \hbar \sum_{\alpha} {\rm Tr} \Big\{ 
\ln (-G_{\alpha}^{-1} ) \Big\}
+ S_{B}[\phi,\phi^{*}]  \nonumber \\
&-& 
\int_{0}^{\hbar\beta} d\tau  \int  d{\bf x}
T_{F} \rho_{\uparrow}({\bf x},\tau) \rho_{\downarrow}({\bf x},\tau) \; ,
\end{eqnarray} 
in terms of the Green's functions
\begin{eqnarray}
&& \hspace*{-0.2in}
G^{-1}_{\alpha}({\bf x},\tau;{\bf x'},\tau')   \nonumber \\
&&= - {1 \over \hbar}
\left\{
\hbar {\partial \over \partial \tau}
- {\hbar^{2} \mbox{\boldmath $\nabla$}^{2} \over 2 m} - \mu_{\alpha}
+ T_{F} \rho_{-\alpha}({\bf x},\tau)
+ T_{\alpha} |\phi({\bf x},\tau)|^{2} \right\}
\delta({\bf x-x'}) \delta(\tau-\tau') \; .
\end{eqnarray} 

Expanding this action around its minimum, by requiring the
linear terms in the fluctuations to be zero, will result 
in the Hartree approximation for the equilibrium densities.  
This is sufficiently accurate because
we are dealing with a fermionic gas in a nonmagnetic phase, 
where the Fock term in the selfenergy is zero, due to the spin-symmetry of the 
action.
It also explains the use of the symbol H(artree) in Eq.~(\ref{eq27}).
Inserting thus into the right-hand side of Eq.~(\ref{eq17})
\begin{eqnarray}
\rho_{\alpha}({\bf x},\tau) & = &
n_{\alpha} +  \rho'_{\alpha}({\bf x},\tau) \; , 
\end{eqnarray}
and
\begin{eqnarray}
\phi({\bf x},\tau) & = & \sqrt{n_{B}} + \phi'({\bf x},\tau) \; , 
\end{eqnarray}
we obtain from the requirement that the linear terms in the fluctuations vanish 
again the Hugenholtz-Pines relation 
\begin{eqnarray}
\label{eq30}
-\mu + T_{B} n_{B} + \sum_{\alpha} T_{\alpha} 
G_{\alpha}^{H}({\bf x},\tau;{\bf x},\tau) & = & 0 \; ,
\end{eqnarray}
and in addition the expected equation for the average spin densities 
\begin{eqnarray}
\label{eq22}
n_{\alpha} = G_{\alpha}^{H}({\bf x},\tau;{\bf x},\tau) \; ,
\end{eqnarray}
where 
\begin{eqnarray}
\label{eq23}
G^{H^{-1}}_{\alpha}({\bf x},\tau;{\bf x'},\tau') = - {1 \over \hbar}
\left\{
\hbar {\partial \over \partial \tau}
- {\hbar^{2} \mbox{\boldmath $\nabla$}^{2} \over 2 m} - \mu_{\alpha}
+ T_{F} \rho_{-\alpha}
+ T_{\alpha} n_{B} \right\}
\delta({\bf x-x'}) \delta(\tau-\tau') \;
\end{eqnarray} 
is the usual fermionic one-particle propagator, or two-point correlation 
function, in the Hartree approximation. 
We, therefore, recognise in Eqs.~(\ref{eq30}), (\ref{eq22}), and (\ref{eq23})
the selfconsistent Hartree equations for the boson and fermion densities
at given chemical potentials $\mu$, $\mu_{\uparrow}$, and $\mu_{\downarrow}$.

To find the theory describing the fluctuations around this equilibrium,
we have to perform the expansion around this minimum up to second order in
the fluctuations $\rho'_{\alpha}$ and $\phi'$.
The poles in the Green's function of the resulting theory
give us the desired dispersion of the collective modes,
for they are also the poles in the linear response
of the densities to an external perturbation.
If we introduce again the vector notation 
\begin{eqnarray}
\bbox{\phi}'_{{\bf k},n} & = &
\left(
\begin{array}{c}
\phi'_{{\bf k},n} \\
\phi'^{*}_{-{\bf k},-n} 
\end{array}
\right) \; ,
\end{eqnarray}
and also 
\begin{eqnarray}
 \bbox{\rho}'_{{\bf k},n} & = &
\left(
\begin{array}{c}
 \rho'_{\uparrow;{\bf k},n} \\
 \rho'_{\downarrow;{\bf k},n}
\end{array}
\right) \; ,
\end{eqnarray} 
the quadratic part of the effective action for the fluctuations 
$\bbox{\rho}'_{\alpha}$ 
and $\bbox{\phi}'$ can be conveniently written in momentum space as
\begin{eqnarray}
\label{eq33}
S^{(2)}[\bbox{\phi}',\bbox{\rho}'] & = &
{1 \over \hbar \beta V} \sum_{{\bf k},n}
{1 \over 2} \left(
\begin{array}{c}
\bbox{\phi}'_{{\bf k},n} \\
 \bbox{\rho}'_{{\bf k},n}
\end{array}
\right)^{\dagger} \cdot
\left(
\begin{array}{cc}
{\bf M}_{\phi\phi}({\bf k},i \omega_{n}) & 
{\bf M}_{\phi\rho}({\bf k},i \omega_{n}) \\
{\bf M}_{\rho\phi}({\bf k},i \omega_{n}) & 
{\bf M}_{\rho\rho}({\bf k},i \omega_{n}) 
\end{array}
\right) \cdot
\left(
\begin{array}{c}
\bbox{\phi}'_{{\bf k},n} \\
 \bbox{\rho}'_{{\bf k},n}
\end{array}
\right) \; ,
\end{eqnarray}
where we have defined the matrices
\begin{eqnarray}
{\bf M}_{\phi\phi}({\bf k},i \omega_{n}) & = & 
\left(
\begin{array}{cc}
-i \hbar \omega_{n} + \epsilon({\bf k})+
\Sigma({\bf k},i \omega_{n})
& \Sigma({\bf k},i \omega_{n}) 
\\
\Sigma({\bf k},i \omega_{n}) & 
i \hbar \omega_{n} + \epsilon({\bf k})+
\Sigma({\bf k},i \omega_{n})
\end{array}
\right) \; ,
\end{eqnarray}
and 
\begin{eqnarray}
{\bf M}_{\rho\rho}({\bf k},i \omega_{n}) & = & 
T_{F}  
\left(
\begin{array}{cc}
T_{F}
\Pi_{\downarrow}({\bf k},i \omega_{n}) 
& - 1 \\
- 1  & 
T_{F} \Pi_{\uparrow}({\bf k},i \omega_{n}) 
\end{array}
\right) \; ,
\end{eqnarray}
and
\begin{eqnarray}
{\bf M}^{T}_{\rho\phi}({\bf k},i \omega_{n}) = 
{\bf M}_{\phi\rho}({\bf k},i \omega_{n}) & = & 
T_{F} \sqrt{n_{B}}
\left(
\begin{array}{cc}
T_{\downarrow} 
\Pi_{\downarrow}({\bf k},i \omega_{n}) & 
T_{\uparrow}
\Pi_{\uparrow}({\bf k},i \omega_{n}) \\
T_{\downarrow}
\Pi_{\downarrow}({\bf k},i \omega_{n}) & 
T_{\uparrow}
\Pi_{\uparrow}({\bf k},i \omega_{n}) 
\end{array}
\right) \; .
\end{eqnarray}
We have also introduced the selfenergy of the Bose gas
\begin{eqnarray}
\Sigma({\bf k},i \omega_{n}) & = &
T_{B} n_{B} + 
\left[ T_{\uparrow}^{2} \Pi_{\uparrow}({\bf k},i \omega_{n})
+ T_{\downarrow}^{2} \Pi_{\downarrow}({\bf k},i \omega_{n}) \right]
n_{B}~,
\end{eqnarray}
and the RPA-bubble, or equivalently, the density-density
response function of the ideal Fermi gas 
\begin{eqnarray}
\label{bubble}
\Pi_{\alpha}({\bf k},i \omega_{n}) =
\int {d {\bf p} \over (2 \pi)^{3}}~ 
{n(\epsilon_{\alpha}({\bf k+p})) - n(\epsilon_{\alpha}({\bf p})) 
\over -i \hbar \omega_{n} + \epsilon_{\alpha}({\bf k+p}) 
- \epsilon_{\alpha}({\bf p})} \; ,
\end{eqnarray}
where the dispersions obey 
$\epsilon_{\alpha}({\bf k})=\hbar^{2} k^{2} / 2m
+ T_{F} \rho_{-\alpha} + T_{\alpha} n_{B} - \mu_{\alpha}$,
and $n(\epsilon)=1/(\exp(\beta \epsilon) + 1)$ is the usual Fermi-Dirac 
distribution function.
Note that $\Sigma({\bf k},i \omega_{n})$ and 
$\Pi_{\alpha}({\bf k},i \omega_{n})$ are invariant under 
the substitution $({\bf k},i \omega_{n}) \rightarrow (-{\bf k},-i \omega_{n})$, 
due to the time-reversal symmetry of the problem. 

In the long-wavelength limit, i.e., for small $k$, 
Eq.~(\ref{bubble}) can be rewritten as
\begin{eqnarray}
\label{eq40}
\Pi_{\alpha}({\bf k},i \omega_{n}) =
{m k_{\alpha} \over 2 \hbar^{2} \pi^{2}}
\int_{0}^{\infty} d\epsilon {\partial n_{\alpha} \over \partial \epsilon}
\sqrt{\epsilon} \left[
1 - {x_{\alpha,n} \over \sqrt{\epsilon}} \arctan{\sqrt{\epsilon} 
\over x_{\alpha,n}}
\right] \; ,
\end{eqnarray}
where $x_{\alpha,n} = m \omega_{n} / \hbar k_{\alpha} k$,
\begin{eqnarray}
n_{\alpha}(\epsilon)=\left[
e^{\beta \left(
\epsilon_{\alpha} \epsilon - 
\mu_{\alpha} + T_{F} \rho_{-\alpha}
+ T_{\alpha} n_{B} \right)}  + 1
\right]^{-1} \; ,
\end{eqnarray}
and we have defined the Fermi energies and wavevectors by
$\epsilon_{\alpha}
= \mu_{\alpha} - T_{F} \rho_{-\alpha} - T_{\alpha} n_{B} 
\equiv \hbar^{2} k_{\alpha}^{2} / 2 m $.
The analytic continuation of Eq.~(\ref{eq40}) to physical energies
$\omega=i \omega_{n}$ reads
\begin{eqnarray}
\label{eq41}
\Pi_{\alpha}({\bf k},\omega) =
{m k_{\alpha} \over 2 \hbar^{2} \pi^{2}}
\int_{0}^{\infty} d\epsilon {\partial n_{\alpha} \over \partial \epsilon}
\left[
1 - {x_{\alpha} \over 2} 
\log \left| {x_{\alpha} + \sqrt{\epsilon} \over x_{\alpha} - \sqrt{\epsilon}} 
\right|
- i {\pi \over 2} |x_{\alpha}| \theta \left(\sqrt{\epsilon} - \left| x_{\alpha} 
\right| \right)
\right] \; ,
\end{eqnarray}
with $x_{\alpha} = m \omega / \hbar k_{\alpha} k$.
We want to find the zero-temperature result and the 
lowest-order corrections in the temperature.
This can be done by means of a Sommerfeld expansion,
which amounts to expanding the real part of the expression 
between square brackets in 
Eq.~(\ref{eq41}) around $\epsilon=1$. Doing so, we find in first instance
\begin{eqnarray}
\label{eq42}
\Pi_{\alpha}({\bf k},\omega) & = &
{m k_{\alpha} \over 2 \hbar^{2} \pi^{2}}
\int_{0}^{\infty} d\epsilon {\partial n_{\alpha} \over \partial \epsilon}
\left\{
1 - {x_{\alpha} \over 2} 
\log \left| {x_{\alpha} + 1 \over x_{\alpha} - 1} \right| \right. 
\nonumber \\
& & - \left[ 1 -
{x_{\alpha}^{2}(x_{\alpha}^{2} - 3) \over (1-x_{\alpha}^{2})^{2}}
\right] {(\epsilon-1)^{2} \over 8} + {\cal O}[(\epsilon-1)^{4}] 
\nonumber \\
& & \left. - i {\pi \over 2} |x_{\alpha}| \theta \left(\sqrt{\epsilon} - 
\left| x_{\alpha} \right| \right) \right\} \; .
\end{eqnarray}
Integrating then over $\epsilon$, we obtain for the real part in lowest order 
\begin{eqnarray}
\label{eq47}
\Pi_{\alpha}^{(0)}({\bf k},\omega) =
- {m k_{\alpha} n_{\alpha}(0) \over 2 \hbar^{2} \pi^{2}}
\left[
1 - {x_{\alpha} \over 2} 
\log \left| {x_{\alpha} + 1 \over x_{\alpha} - 1} \right|
\right] \; ,
\end{eqnarray}
and for the imaginary part exactly 
\begin{eqnarray}
\label{eq48}
{\rm Im} \left[ \Pi_{\alpha}({\bf k},\omega) \right] =
{m k_{\alpha} n_{\alpha}(0) \over 2 \hbar^{2} \pi^{2}}
{\pi \over 2 n_{\alpha}(0)} |x_{\alpha}| 
n_{\alpha} \left( |x_{\alpha}|^{2} \right) \; .
\end{eqnarray}
At zero temperature Eqs.~(\ref{eq47}) and (\ref{eq48})
reduce to the well-known result for the zero-temperature
RPA bubble \cite{fetter and walecka}. Note that the imaginairy
part in Eq.~(\ref{eq48}) is valid for all temperatures and cannot simply be 
expanded as a power series in $k_{B} T / \epsilon_{F}$.
This is not true for the real part of $\Pi_{\alpha}({\bf k},\omega)$ and the 
lowest-order temperature correction is given by
\begin{eqnarray}
\label{eq49a}
\Pi_{\alpha}^{(2)}({\bf k},\omega) & = &
- {m k_{\alpha} \over 16 \hbar^{2} \pi^{2}}
\left[ 1 -
{x_{\alpha}^{2}(x_{\alpha}^{2}-3) \over (1-x_{\alpha}^{2})^{2}} 
\right] 
\left( {k_{B} T \over \epsilon_{\alpha}} \right)^{2} 
{\pi^{2} \over 3} \; .
\end{eqnarray} 

To find the long-wavelength dispersion, i.e., $\omega = c k$ or
equivalently $x_{\alpha} = m c / \hbar k_{\alpha}$, of the 
collective modes of the mixture, we require that  
the determinant of the fluctuation matrix  
in Eq.~(\ref{eq33}) equals zero. Calculating the determinant, 
the terms of
${\cal O}(k^{0})$ drop out, indicating that the collective 
excitations are indeed gapless as assumed by our
ansatz $\omega=c k$. Leaving out terms that 
are of ${\cal O}(k^{4})$, because they do not affect the
linear part of the dispersion relation, and dividing out an overall
factor of $k^{2}$, we ultimately find the result
\begin{eqnarray}
\label{eq50}
\left[ T_{F}^{2} 
\Pi_{\uparrow}(c) \Pi_{\downarrow}(c) - 1 \right]
\left[ -mc^{2} + \Sigma(c) \right] 
\hspace*{0.5in} &&
\nonumber \\
-T_{F} n_{B} \left[
T_{F} \Pi_{\uparrow}(c) T_{\downarrow} 
\Pi_{\downarrow}(c) -
T_{\uparrow} \Pi_{\uparrow}(c)
\right] T_{\downarrow} \Pi_{\downarrow}(c)
& & 
\nonumber \\
-T_{F} n_{B} \left[ 
T_{F} \Pi_{\downarrow}(c) T_{\uparrow} 
\Pi_{\uparrow}(c) -
T_{\downarrow} \Pi_{\downarrow}(c)
\right] T_{\uparrow} \Pi_{\uparrow}(c)
&=& 0 \; .
\end{eqnarray}
Here, we have for convenience introduced the shorthand notation 
$\Pi_{\alpha}(c)=\Pi_{\alpha}({\bf k},ck)$.
If we put $T_{\uparrow}=T_{\downarrow}=0$, this simply gives
\begin{eqnarray}
\left[ T_{F}^{2} 
\Pi_{\uparrow}(c) \Pi_{\downarrow}(c) - 1 \right]
\left[ -mc^{2} + \Sigma(c) \right] 
& = & 0 \; .
\end{eqnarray}
The first factor describes the collective modes of the 
Fermi gas and the second factor describes the Bogoliubov modes
of the condesate, which are of course decoupled in this case. For small values 
of $T_{\uparrow}$ and $T_{\downarrow}$ these 
modes also exist, but the dispersion is changed.
The lowest-order correction 
in the boson-fermion scattering lengths
$a_{\alpha}$ to the Bogoliubov speed of sound 
$c_{0} = \sqrt{T_{B} n_{B} / m}$ is determined by
\begin{eqnarray}
\label{eq49}
m c_{1}^{2} = \Sigma(c_{0}) 
& - &
{T_{F} n_{B} \left[
T_{F} \Pi_{\uparrow}(c_{0}) 
T_{\downarrow} \Pi_{\downarrow}(c_{0}) -
T_{\uparrow} \Pi_{\uparrow}(c_{0})
\right] T_{\downarrow} \Pi_{\downarrow}(c_{0})
\over 
\left[ T_{F}^{2} 
\Pi_{\uparrow}(c_{0}) \Pi_{\downarrow}(c_{0}) - 1 \right]}
\nonumber \\ 
& - &
{T_{F} n_{B} \left[ 
T_{F} \Pi_{\downarrow}(c_{0}) T_{\uparrow} 
\Pi_{\uparrow}(c_{0}) -
T_{\downarrow} \Pi_{\downarrow}(c_{0})
\right] T_{\uparrow} \Pi_{\uparrow}(c_{0})
\over 
\left[ T_{F}^{2} 
\Pi_{\uparrow}(c_{0}) \Pi_{\downarrow}(c_{0}) - 1 \right]} \; .
\end{eqnarray}
To find the full solution we can just
iterate Eq.~(\ref{eq49}) and the result converges 
rapidly to a solution of Eq.~(\ref{eq50}). 
Note that Eq.~(\ref{eq49}) has an imaginary part
and therefore also describes the damping of the 
Bogoliubov mode. Under the circumstances that we have studied in 
Secs.~\ref{sec2a} and \ref{sec2b} the correction on the speed of sound due to 
the presence of the Fermi gas is small and at most about 5\% of the uncoupled 
Bogoliubov result. This is important for our purposes, because it shows that for 
the calculation of the effective fermion-fermion interaction, we do not need to 
selfconsistently include the effect of the Fermi gas on the density fluctuations 
of the condensate. 

To find the other propagating
solutions of Eq.~(\ref{eq50}) we need to be more careful.
For clarity, we first again treat the case where 
$T_{\uparrow}=T_{\downarrow}=0$. The zero-sound modes are now the solutions of 
\begin{eqnarray}
\label{eq53}
\left[ T_{F}^{2} 
\Pi_{\uparrow}(c) \Pi_{\downarrow}(c) - 1 \right] = 0 \; .
\end{eqnarray}
The product $\Pi_{\uparrow}(c) \Pi_{\downarrow}(c)$ diverges 
logarithmically for two value of $c$, i.e., for the Fermi velocities
$c_{\alpha}=\hbar k_{\alpha} / m$. In principle
there are therefore four solutions to this equation, which can be found by 
expanding around either $c_{\uparrow}$ or $c_{\downarrow}$.
Without loss of generality, we assume that 
$c_{\uparrow} > c_{\downarrow}$.
To find then the solution that is not overdamped,
we have to expand around $c_{\uparrow}$ and find the
solution which is slightly bigger than $c_{\uparrow}$, i.e.,
$c=c_{\uparrow}+\delta c$ with $\delta c > 0$. In this way we make sure that
the imaginary parts of both 
$\Pi_{\uparrow}(c)$ and $\Pi_{\downarrow}(c)$ are equal to zero at zero
temperature.
Experimentally we are always in the weak-coupling limit, which implies that
$T_{F} \ll \hbar^{2} / m k_{\uparrow}$.
In this case the zero-temperature expressions for $\Pi_{\uparrow}(c)$ and 
$\Pi_{\downarrow}(c)$ become explicitly
\begin{eqnarray}
\Pi_{\uparrow}^{0}(c_{\uparrow}+\delta c) =
- {m k_{\uparrow} \over 2 \hbar^{2} \pi^{2}}
\left[
1 - {1\over 2} 
\log {2 c_{\uparrow} \over \delta c} 
\right] + {\cal O}(\delta c)
\; ,
\end{eqnarray}
and
\begin{eqnarray}
\Pi_{\downarrow}^{0}(c_{\uparrow}+\delta c) =
- {m k_{\downarrow} \over 2 \hbar^{2} \pi^{2}}
\left[
1 - 
{1 \over 2}
\left( 
{c_{\uparrow} \over c_{\downarrow}}
\right) 
\log { \left( {c_{\uparrow} \over c_{\downarrow}}  \right) + 1\over 
\left( {c_{\uparrow} \over c_{\downarrow}} \right) +
{\delta c \over c_{\downarrow}} - 1}
\right] 
+ {\cal O}(\delta c) \; ,
\end{eqnarray}
respectively. In the limit that 
$(c_{\uparrow} - c_{\downarrow}) \ll \delta c$
we thus get
\begin{eqnarray}
\label{eq56}
\delta c = 2 c_{\uparrow} \exp \left(-{4 \pi^{2} \hbar^{2} \over m 
k_{\uparrow} T_{F}} - 2 \right) \; ,
\end{eqnarray}
where we have made use of the fact that in this limit
$k_{\downarrow}-k_{\uparrow}={\cal O}(\delta c)$.
Note that our result differs from that of Fetter and Walecka 
\cite{fetter and walecka}.
They use an approximation that does not 
obey the Pauli exclusion principle, because there is s-wave 
scattering between particles that are in the same spin-state.
As a result, their mean-field energy for an atom in the spin state 
$|\alpha\rangle$ is $T_{F} \sum_{\alpha} n_{\alpha}$
instead of $T_{F} n_{-\alpha}$. For $n_{\uparrow}=n_{\downarrow}$ 
this effectively means a factor of two reduction of the interaction
strength. In the limit that 
$(c_{\uparrow} - c_{\downarrow}) \gg \delta c$
we find at zero temperature
\begin{eqnarray}
\label{eq57}
\delta c = 2 c_{\uparrow} \exp \left(-{4 \pi^{2} \hbar^{2} \over m 
k_{\uparrow} T_{F}^{2} \Pi_{\downarrow}^{0}(c_{\uparrow})} - 2 \right) \; .
\end{eqnarray}

Let us now consider the effect of the presence of the Bose condensate and do the 
same calculation for the mixture of the Bose condensed
gas with the two-component fermion gas. 
In the limit that 
$(c_{\uparrow} - c_{\downarrow}) \ll \delta c$ we again find
\begin{eqnarray}
\delta c = 2 c_{\uparrow} \exp \left(-{4 \pi^{2} \hbar^{2} \over m 
k_{\uparrow} T_{F}} - 2 \right) \; .
\end{eqnarray}
This is the same result as in Eq.~(\ref{eq56}) where the fermionic and 
the Bose condensed gas are decoupled, and is due to the fact that 
in this limit the zero-sound mode is a pure spin wave.
The density profile is therefore constant and,
since the fermionic and the Bose condensed
gas couple only through the density fluctuations, 
there is no effect of the presence
of the Bose gas. On the other hand, in the limit that 
$(c_{\uparrow} - c_{\downarrow}) \gg \delta c$ we now obtain 
\begin{eqnarray}
\label{eq59}
\delta c = 2 c_{\uparrow} \exp \left(-{4 \pi^{2} \hbar^{2} \over m 
k_{\uparrow} T_{F} \chi} - 2 \right) \; ,
\end{eqnarray}
where we have defined the quantity
\begin{eqnarray}
\chi = {T_F(-m c_{\uparrow}^{2} + T_{B} n_{B} + T_{\downarrow}^{2} 
\Pi^{0}_{\downarrow}(c_{\uparrow}) n_{B}) \over
T_{F}^2 \Pi^{0}_{\downarrow}(c_{\uparrow})
(-m c_{\uparrow}^{2} + T_{B} n_{B}) -
T_{\uparrow}^{2} n_{B} +
2 T_{F} T_{\downarrow} T_{\uparrow} 
\Pi^{0}_{\downarrow}(c_{\uparrow}) n_{B}} \; .
\end{eqnarray}
Hence, Eq.~(\ref{eq59}) reduces to Eq.~(\ref{eq57}) if we take   
$T_{\uparrow}=T_{\downarrow}=0$, as it should. The same is of course true if 
$n_B = 0$. Note that Eq.~(\ref{eq49a}) offers the opportunity to obtain also the 
small nonzero-temperature corrections to the above results, if this turns out to 
be necessary for a particular application.

\section{CONCLUSIONS}
\label{sec2d}
We have calculated the effect of phonon exchange on the
scattering length for two fermions with different spins.
Under the right circumstances this effect can be quite large 
and is then a possible way to experimentally tune the 
fermion-fermion scattering length. This may in particular be useful for 
the achievement of a BCS transition in a
mixture of $^{40}$K and $^{39}$K, although it appears that a more 
precise determination of the various scattering lengths involved in this
case is necessary to make sure of this.
In addition, we have analyzed the stability and
the mode structure of the Bose-Fermi mixture.
Our results in the latter case, 
which are valid for an arbitrary ratio of the densities of the
three components, agree in limiting cases with expressions obtained by other 
authors \cite{ho3}. 

\section*{ACKNOWLEDGMENTS}
We would like to thank C.J. Pethick and E. Braaten for 
usefull comments.

\begin{figure}[h]
\vspace{0.25 cm}
\caption{
\label{fig1}
The phase shift $\delta_{0}$ as a function of the momentum $k$, for a mixture
of $^{40}$K and $^{87}$Rb. The condensate density is
(1) $n_{B}=1 \times 10^{10}$ cm$^{-3}$, (2) $n_{B}=1 \times 10^{13}$ cm$^{-3}$, 
and (3) $n_{B}=1 \times 10^{14}$ cm$^{-3}$.}
\end{figure}

\begin{figure}[h]
\vspace{0.25 cm}
\caption{
\label{fig2}
The phase shift $\delta_{0}$ as a function of the momentum $k$, for a mixture
of $^{6}$Li and $^{87}$Rb.  The condensate density is
(1) $n_{B}=1 \times 10^{10}$ cm$^{-3}$, (2) $n_{B}=1 \times 10^{13}$ cm$^{-3}$, 
and (3) $n_{B}=1 \times 10^{14}$ cm$^{-3}$.} 
\end{figure}

\begin{figure}[h]
\vspace{0.25 cm}
\caption{
\label{fig3}
The phase shift $\delta_{0}$ as a function of the momentum $k$, for a mixture
of $^{40}$K and $^{39}$K. The condensate density is
(1) $n_{B}=4 \times 10^{12}$ cm$^{-3}$,
(2) $n_{B}=5 \times 10^{12}$ cm$^{-3}$,
(3) $n_{B}=6 \times 10^{12}$ cm$^{-3}$,
(4) $n_{B}=7 \times 10^{12}$ cm$^{-3}$,
(5) $n_{B}=8 \times 10^{12}$ cm$^{-3}$. }
\end{figure}

\begin{figure}[h]
\vspace{0.25 cm}
\caption{
\label{fig4}
The scattering length $a_F^{\rm eff}$
as a function of the density of condensed atoms $n_{B}$, for a mixture
of $^{40}$K and $^{39}$K.}
\end{figure}

\end{document}